\documentclass[aps,prd,twocolumn,10pt,superscriptaddress,nofootinbib,longbibliography]{revtex4-1}

\usepackage{amssymb}
\usepackage{graphicx}
\usepackage{amsmath}
\usepackage{hyperref}
\usepackage{multirow}
\usepackage{setspace}
\usepackage{verbatim}
\usepackage{float}
\usepackage{color}
\usepackage{ulem}
\usepackage[utf8]{inputenc}
\ifdefined\DeclareUnicodeCharacter
  \DeclareUnicodeCharacter{2212}{\ensuremath{-}}
\fi
\usepackage[table,xcdraw]{xcolor}
\usepackage{makecell}
\usepackage{url}
\usepackage{bm}

\begin{document}

\title{Nonminimally coupled quintessence with sign-switching interaction}

\author{Jia-Qi Wang}
\email{wangjiaqi@itp.ac.cn}
\affiliation{Institute of Theoretical Physics, Chinese Academy of Sciences (CAS), Beijing 100190, China}
\affiliation{University of Chinese Academy of Sciences (UCAS), Beijing 100049, China}

\author{Rong-Gen Cai}
\email{caironggen@nbu.edu.cn}
\affiliation{Institute of Fundamental Physics and Quantum Technology, Ningbo University, Ningbo, 315211, China}

\author{Zong-Kuan Guo}
\email{guozk@itp.ac.cn}
\affiliation{Institute of Theoretical Physics, Chinese Academy of Sciences (CAS), Beijing 100190, China}
\affiliation{University of Chinese Academy of Sciences (UCAS), Beijing 100049, China}
\affiliation{School of Fundamental Physics and Mathematical Sciences, Hangzhou Institute for Advanced Study, University of Chinese Academy of Sciences, Hangzhou 310024, China}

\author{Yun-He Li}
\email{liyunhe@mail.neu.edu.cn}
\affiliation{Department of Physics, College of Sciences, Northeastern University, Shenyang 110819, China}

\author{Shao-Jiang Wang}
\email{schwang@itp.ac.cn}
\affiliation{Institute of Theoretical Physics, Chinese Academy of Sciences (CAS), Beijing 100190, China}
\affiliation{Asia Pacific Center for Theoretical Physics (APCTP), Pohang 37673, Korea}

\author{Xin Zhang}
\email{zhangxin@mail.neu.edu.cn}
\affiliation{Department of Physics, College of Sciences, Northeastern University, Shenyang 110819, China}
\affiliation{Key Laboratory of Cosmology and Astrophysics (Liaoning), Northeastern University, Shenyang 110819, China}
\affiliation{Key Laboratory of Data Analytics and Optimization for Smart Industry (Ministry of Education), Northeastern University, Shenyang 110819, China}
\affiliation{National Frontiers Science Center for Industrial Intelligence and Systems Optimization, Northeastern University, Shenyang 110819, China}

\begin{abstract}
We propose a new nonminimally coupled quintessence model to account for the late-time dark energy dynamics indicated by recent Dark Energy Spectroscopic Instrument (DESI) measurements.
Within this framework, the quintessence density begins to decrease only when it starts to dominate the Universe, which naturally accounts for the late-time onset of dark energy weakening.
The coupling also induces a sign change in the effective energy transfer between dark matter and dark energy during cosmic evolution. While the scalar field itself remains canonical and never crosses the phantom divide, the modified evolution of the dark matter density gives rise to an effective crossing behavior in the observationally inferred dark energy sector. Compared with both $\Lambda\mathrm{CDM}$ and $w_0w_a\mathrm{CDM}$ models, our model is favored more strongly by current cosmological data. This work may provide a promising avenue for understanding the observational late-time weakening of dark energy and the origin of its dynamics.

\end{abstract}
\maketitle

\section{Introduction}\label{sec:intro}

Over the past decade, precision cosmology has entered an era in which multiple independent probes can map the expansion history of the Universe with unprecedented accuracy. The measurements of cosmic microwave background (CMB) by Planck~\cite{Planck:2018vyg}, baryon acoustic oscillation (BAO) measurements from eBOSS~\cite{eBOSS:2020yzd} as well as the Dark Energy Spectroscopic Instrument (DESI)~\cite{DESI:2024mwx}, and large type Ia supernova (SN) compilations including PantheonPlus~\cite{Brout:2022vxf} and the recent Dark Energy Survey (DES) five-year supernova analysis~\cite{DES:2024jxu} together provide remarkably stringent tests of the standard cosmological framework. Within this observational landscape, the $\Lambda$-cold-dark-matter ($\Lambda$CDM) model has achieved extraordinary success and remains the benchmark description of modern cosmology~\cite{Weinberg:1988cp,Frieman:2008sn}. However, several persistent challenges remain. The cosmological constant and coincidence problems continue to motivate alternatives to a strictly constant vacuum energy~\cite{Peebles:2002gy,Copeland:2006wr}, while the discrepancy between the early Universe inference of the Hubble constant ($H_{0}=67.49\pm0.53\, {\rm km/s/Mpc}$)~\cite{Planck:2018vyg,ACT:2020gnv,SPT-3G:2021wgf} and late-Universe distance ladder determinations ($H_0=73.01\pm 1.04 \, \mathrm{km/s/Mpc}$)~\cite{Riess:2016jrr,Riess:2018uxu, Riess:2019cxk,Riess:2021jrx} has not yet found a generally accepted resolution~\cite{DiValentino:2021izs,Poggiani:2025hbe,Perivolaropoulos:2021jda,Hu:2023jqc,Kamionkowski:2022pkx}. At the same time, recent combinations of BAO and supernova data have revived interest in the possibility that cosmic acceleration may be weakening at late times rather than being driven by a cosmological constant~\cite{DESI:2024mwx,DES:2024jxu}.

These persistent tensions have strengthened the possibility that dark energy (DE) may not be a strict cosmological constant, but instead may possess nontrivial late-time dynamics~\cite{Peebles:2002gy,Copeland:2006wr}. A widely used phenomenological framework for testing this possibility is the Chevallier-Polarski-Linder (CPL) parametrization~\cite{Chevallier:2000qy,Linder:2002et}, in which the equation of state (EoS) of DE is written as $w(a)=w_{0}+w_{a}(1-a)$. In the recent DESI analysis~\cite{DESI:2025fii,DESI:2025zgx}, this parametrization provides a substantially better fit than the $\Lambda\mathrm{CDM}$ model. Within spatially flat $w_{0}w_{a}$CDM, the preference for evolving DE reaches about $2.5\sigma$, $3.5\sigma$, and $3.9\sigma$ for the combined datasets with Pantheon+, Union3, and DES-SN5YR supernova data, respectively~\cite{DESI:2024mwx}. Although the evidence for evolving DE can be affected by the choice and calibration of supernova samples~\cite{Efstathiou:2024xcq,DES:2025tir,Huang:2025som,DES:2025sig,Colgain:2025nzf,Montefalcone:2026iga} (see also Ref.~\cite{Ling:2025lmw} for a model-independent illustration), the Planck+DESI combination without supernova data already shows a preference for the CPL extension over $\Lambda\mathrm{CDM}$ at about $2.6\sigma$~\cite{DESI:2024mwx}. Moreover, current observations prefer a crossing behavior of DE in the $w_0w_a\mathrm{CDM}$ model, namely a crossing of the phantom divide from $w<-1$ to $w>-1$ during cosmic evolution. Equivalently, over part of cosmic history, the inferred DE density increases as the Universe expands~\cite{Gonzalez-Fuentes:2026rgu}. This behavior cannot be realized by a single canonical quintessence field with a regular Lagrangian~\cite{Vikman:2004dc}.

Therefore, physically motivated models in which a canonical quintessence field is nonminimally coupled either to gravity or to dark matter (DM) have attracted renewed attention, building on the longstanding literature on coupled DE~\cite{Guo:2004vg,Gomez-Valent:2020mqn,Goh:2022gxo,Guo:2004xx,Guo:2007zk,Yang:2015tzc,Wetterich:1994bg,Amendola:1999er,Bolotin:2013jpa,Ye:2024ywg,Samanta:2025oqz,Adam:2025kve,Wolf:2025jed,Wang:2025znm,SanchezLopez:2025uzw}. Recent analyses indicate that such scenarios can accommodate the late-time evolution of the dark sector with evidence for nonminimal coupling exceeding $2\sigma$ level~\cite{Wolf:2025jed,Wang:2025znm,SanchezLopez:2025uzw,Li:2026xaz}. However, the preferred direction of energy transfer remains unsettled, since it changes as the interaction ansatz is varied. Nonparametric reconstructions of coupled fluid models further suggest that the coupling direction may reverse at late times~\cite{Li:2024qso,Cai:2009ht,Li:2025ula,Li:2025owk,Yang:2025uyv,Wu:2025vrl,Sabogal:2025mkp}. Clarifying these issues, together with understanding why the scalar field begins to evolve appreciably only at late times, makes the late-time behavior of DE in concrete coupled quintessence models a particularly worthwhile subject of investigation.

In this work, we propose a dark-sector extension of $\Lambda\mathrm{CDM}$, denoted $\beta\varphi\mathrm{CDM}$, in which a canonical scalar field is coupled to DM through a quadratic conformal factor and evolves under an exponential potential~\cite{Amendola:1999qq,Amendola:1999er,vanDeBruck:2019vzd,Kase:2019veo,Zhang:2005rg,Lyu:2025nsd}. This setup allows the field to remain effectively frozen throughout most of the early expansion history and to thaw only once DE becomes dominant, thereby providing a natural explanation for why DE begins to weaken only at late times~\cite{Wetterich:1994bg,Amendola:1999er,Bolotin:2013jpa,Ye:2024ywg,Wolf:2025jed,Wang:2025znm,Samanta:2025oqz,Adam:2025kve,SanchezLopez:2025uzw,Li:2026xaz}. As a result, the $\beta\varphi\mathrm{CDM}$ model is preferred by current measurements over both $\Lambda\mathrm{CDM}$ and $w_{0}w_{a}\mathrm{CDM}$ models. The nonstandard evolution of DM induced by the coupling gives rise to an effective crossing behavior in the observationally inferred DE sector~\cite{Wang:2025znm,Liu:2025bss}. Furthermore, the $H_{0}$ and $S_{8}$ tensions are at least not worsened in this setup, in line with the broader tendency of interacting dark-sector models to alleviate one or both of these discrepancies~\cite{Kase:2019veo,Shah:2024rme}. Since the coupling is confined to the dark sector, the model is also not directly excluded by laboratory searches for fifth forces, and therefore needs no external theory to screen the coupling effect~\cite{Koyama:2009gd,vanDeBruck:2019vzd}.

This paper is organized as follows. In Sec.~\ref{sec:model}, we present the $\beta\varphi\mathrm{CDM}$ model and its background dynamics. In Sec.~\ref{sec:method}, we describe the methodology and datasets adopted in the analysis. We report the cosmological constraints and Bayes evidence for the model in Sec.~\ref{sec:constraint}, and discuss the resulting dark-energy evolution, the effective crossing behavior, and the implications for the Hubble tension in Sec.~\ref{sec:discussion}. We conclude in Sec.~\ref{sec:conclusion}.

\section{Model Setup}\label{sec:model}

We construct the $\beta\varphi\mathrm{CDM}$ model by modifying the Einstein-Hilbert and standard-model (SM) actions as
\begin{equation}
    S=S_\mathrm{GR}+S_\mathrm{SM}+S_\mathrm{Dark},
\end{equation}
where the gravitational sector $S_\mathrm{GR}$ and the particle-physics SM sector $S_\mathrm{SM}$ remain unchanged. Relative to $\Lambda\mathrm{CDM}$, all extensions are confined to the dark sector $S_\mathrm{Dark}$. In the present $\beta\varphi\mathrm{CDM}$ framework, the dark-sector action can still be decomposed into a modified DM part, $S_\mathrm{DM}$, and a DE part, $S_\varphi$, described by a canonical scalar field. The actions are given by
\begin{align}
    S_{\varphi} &= \int d^4x\,\sqrt{-g}\left[-\frac{1}{2}g^{\mu\nu}\partial_\mu\varphi\,\partial_\nu\varphi - V(\varphi)\right],\label{eq:scalar_action}\\
    S_\mathrm{DM} &= \int d^4x\,\sqrt{-\tilde{g}}\,\mathcal{L}_\mathrm{CDM}\left[\psi;\tilde{g}_{\mu\nu}\right],\label{eq:DM_action}
\end{align}
where $\mathcal{L}_\mathrm{CDM}\left[\psi;\tilde{g}_{\mu\nu}\right]$ denotes the CDM Lagrangian, with the metric conformally transformed to $\tilde{g}_{\mu\nu}$~\cite{Amendola:1999er}. In nonminimally coupled quintessence models, such a conformal transformation can be written as $\tilde{g}_{\mu\nu}=\mathcal{A}^2(\varphi)\,g_{\mu\nu}$, and the corresponding equations of motion become~\cite{Amendola:1999er,Wetterich:1994bg, Khoury:2003rn,Upadhye:2012vh}
\begin{align}
    \dot\rho_\varphi + 3H(1+w_\varphi)\rho_\varphi &= -\frac{\mathcal{A}^\prime(\varphi)}{\mathcal{A}(\varphi)}\dot{\varphi}\rho_\mathrm{DM}, \label{eq:EOM_phi}\\
    \dot\rho_{\rm DM} + 3H\rho_{\rm DM} &= +\frac{\mathcal{A}^\prime(\varphi)}{\mathcal{A}(\varphi)}\dot{\varphi}\rho_\mathrm{DM},\label{eq:EOM_DM}
\end{align}
where the CDM equation-of-state (EoS) parameter has been fixed to $w_\mathrm{DM}=0$.

The nonminimal coupling modifies the evolution of both DM and DE. Direct integration of Eq.~\eqref{eq:EOM_DM} yields
\begin{equation}
    \rho_\mathrm{DM}\propto \mathcal{A}(\varphi)a^{-3}.\label{eq:Evo_DM}
\end{equation}
Compared with $\Lambda\mathrm{CDM}$, the effect of the nonminimal coupling is simply to introduce an additional conformal factor $\mathcal{A}(\varphi)$ into the evolution of $\rho_\mathrm{DM}$. This conformal factor also appears in the Klein-Gordon equation for quintessence, such that the evolution of the scalar field $\varphi$ can be regarded as being governed by an effective potential,
\begin{equation}
    V_\mathrm{eff}^{\prime}(\varphi,\rho_\mathrm{DM})=V^\prime(\varphi)+\frac{\mathcal{A}^\prime(\varphi)}{\mathcal{A}(\varphi)}\rho_\mathrm{DM}.\label{eq:Veff}
\end{equation}
To proceed, we adopt a simple yet representative choice of a quadratic coupling function (see also Refs.~\cite{Pitrou:2023swx,Uzan:2023dsk}) and an exponential potential in the $\beta\varphi\mathrm{CDM}$ model,
\begin{equation}
    \mathcal{A}(\varphi)=1+\frac{1}{2}\beta\varphi^2,
    \qquad
    V(\varphi)=V_0\exp(\alpha\varphi).
    \label{eq:AV_choice}
\end{equation}
Taking into account the normalization condition for the total cosmic energy budget, the coefficient $V_0$ in the potential is not a free parameter, but is fixed by the boundary condition at $z=0$~\cite{Wang:2025znm}. Therefore, the $\beta\varphi\mathrm{CDM}$ model introduces three additional free parameters, $\alpha$, $\beta$, and the initial value of the scalar field, $\varphi_\mathrm{ini}$.

\begin{figure}[hptb]
    \centering
    \includegraphics[width=\linewidth]{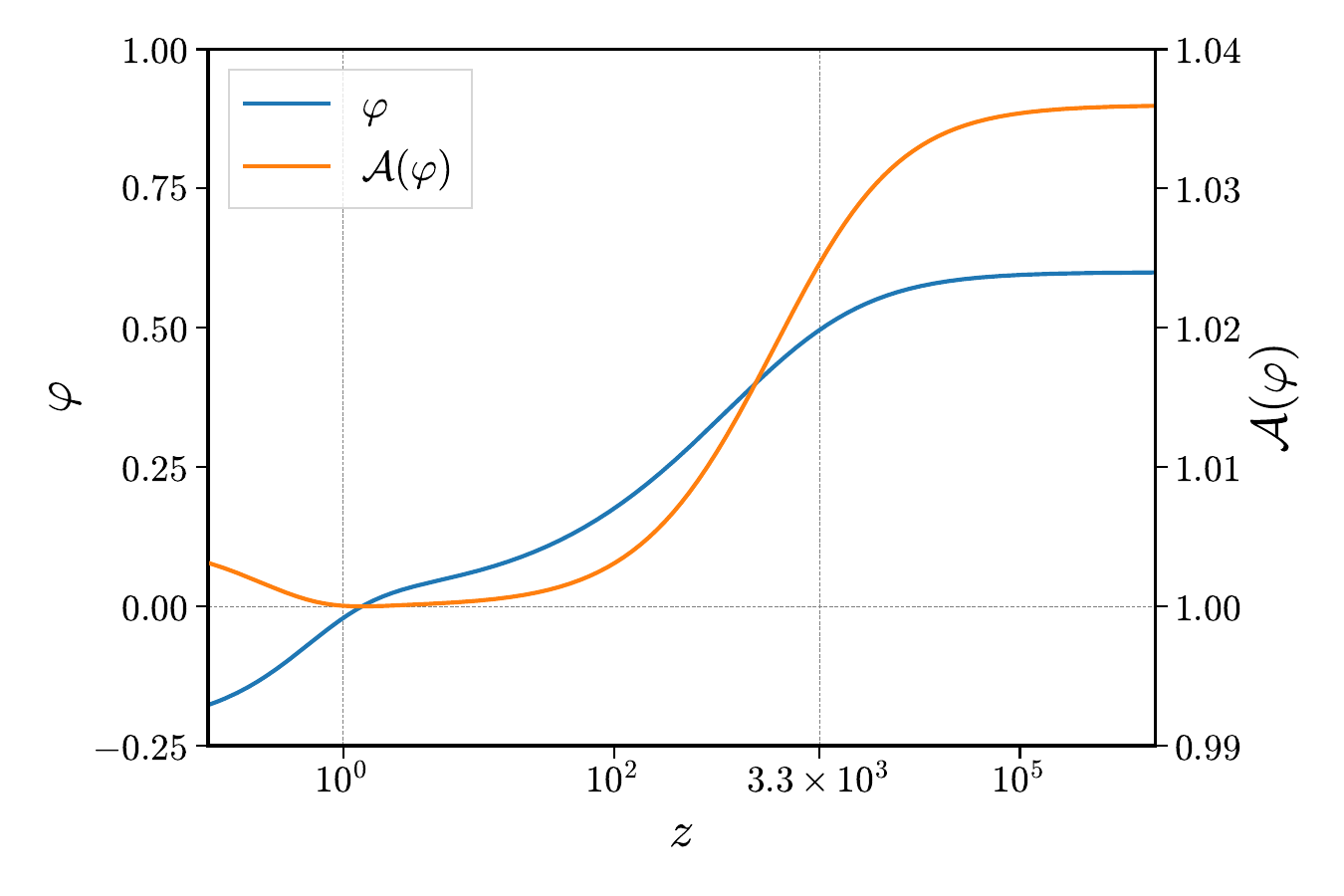}
    \caption{Background evolution of the scalar field $\varphi$ (blue line) and the coupling factor $\mathcal{A}(\varphi)$ (orange line) as a function of redshift. The representative parameters are $\{\alpha,\beta,\varphi_{\rm ini}\}=\{0.8,0.2,0.6\}$.}
    \label{fig:phi_evolution}
\end{figure}

The solution to Eq.~\eqref{eq:EOM_phi} is shown in Fig.~\ref{fig:phi_evolution}. In particular, the evolution of the field $\varphi$ can be divided into the following three stages. (i) During the radiation-dominated era, the evolution of quintessence is dominated by Hubble friction. The attractor solution of Eq.~\eqref{eq:EOM_phi} indicates that the quintessence field is frozen at the initial value $\varphi_\mathrm{ini}$. (ii) As matter begins to dominate the Universe, the scalar field gradually thaws and evolves toward $\varphi=0$ under the dominance of the coupling term. (iii) Since the matter density decays as $a^{-3}$, the potential $V(\varphi)$ becomes dynamically important in the late Universe. The analytic derivation of this attractor behavior is given in Appendix~\ref{app:attractor}.

We also show in Fig.~\ref{fig:Veff} the effective potential $V_\mathrm{eff}$ defined by Eq.~\eqref{eq:Veff} for $z<5$. Intriguingly, as the quintessence density exceeds the matter density, the minimum of the effective potential gradually shifts away from $\varphi=0$ toward negative values. As a result, $\varphi$ begins to thaw from $\varphi=0$ and continues to decrease. Although the scalar field always rolls toward the minimum of $V_\mathrm{eff}(\varphi)$, the trend of the conformal factor $\mathcal{A}(\varphi)$ \textit{reverses} as $\varphi$ crosses zero, as shown in Fig.~\ref{fig:phi_evolution}. Such a nonminimal coupling allows the effective energy transfer to proceed from DM to DE in the early Universe, but from DE to DM at late times. This behavior is consistent with the coupling trend reconstructed nonparametrically in Ref.~\cite{Li:2025ula,Li:2025owk}.

The key feature of the $\beta\varphi\mathrm{CDM}$ model is that $\mathcal{A}(\varphi)$ is an even function, whereas $V(\varphi)$ is monotonic. Other particle-physics-motivated models satisfying these conditions might as well be able to achieve the same cosmological behavior realized in this toy model.

\begin{figure}[t]
    \centering
    \includegraphics[width=\linewidth]{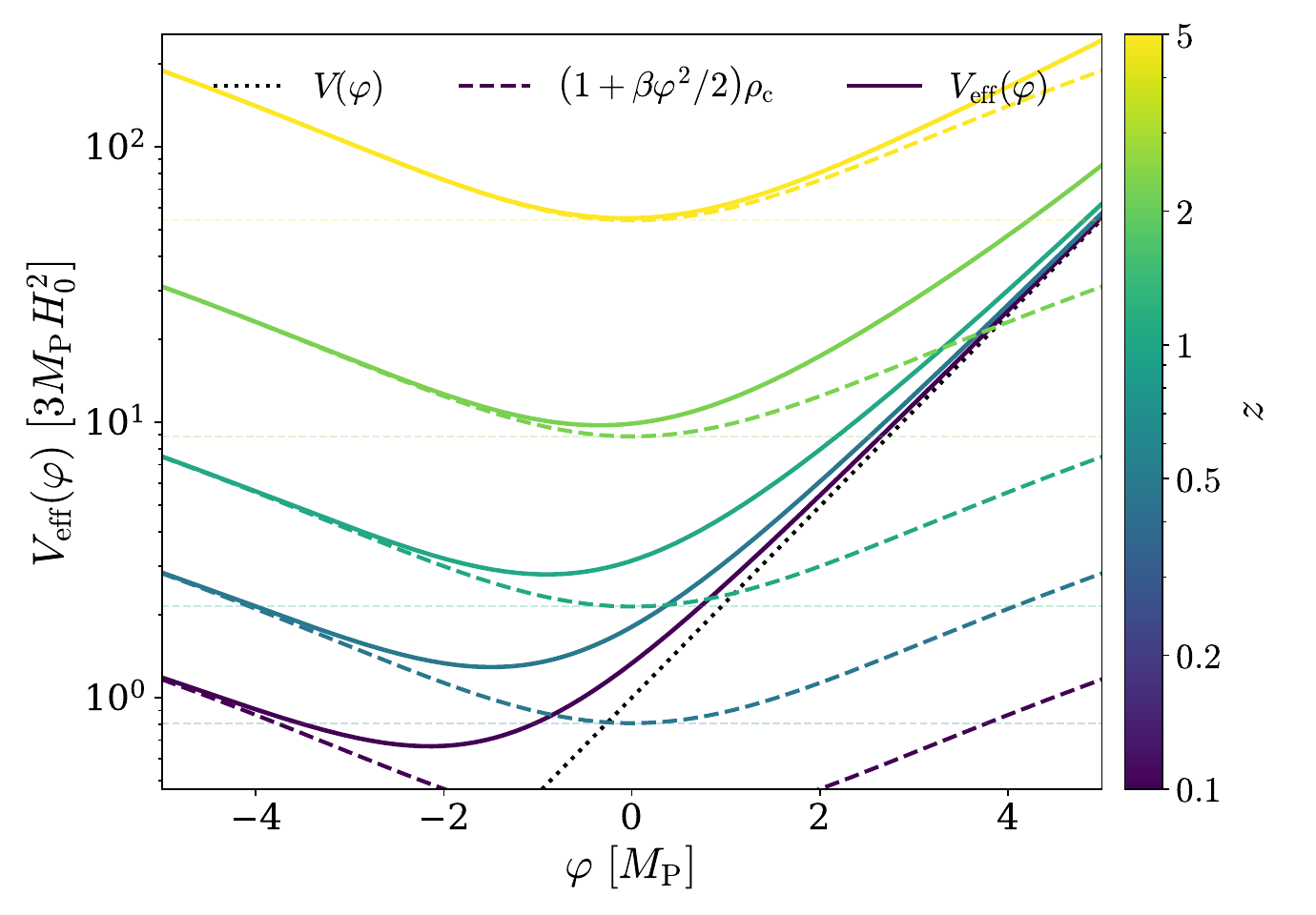}
    \caption{Effective potential $V_{\rm eff}(\varphi)$ defined by Eq.~\eqref{eq:Veff} in the late Universe ($0<z\leq 5$), in which $\{\alpha,\beta\}=\{0.8,0.2\}$ }
    \label{fig:Veff}
\end{figure}

\section{Methodology and Data}\label{sec:method}

\begin{table}[b]
\centering
\caption{Uniform priors for all cosmological and model parameters used in the nested-sampling analysis.}
\label{tab:priors}
\renewcommand{\arraystretch}{1.25}

\newcommand{\wA}{1.75cm}
\newcommand{\wB}{2.75cm}
\newcommand{\wC}{1.0cm}
\newcommand{\wD}{2.5cm}

\begin{tabular}{l c|l c}
\hline\hline
\multicolumn{2}{c|}{Cosmological Parameters} & \multicolumn{2}{c}{Model Parameters} \\
\hline
\makebox[\wA][l]{$\Omega_b h^2$} & \makebox[\wB][c]{$U[0.020,\,0.025]$}
& \makebox[\wC][c]{$w_0$} & \makebox[\wD][c]{$U[-3,\,1]$} \\

\makebox[\wA][l]{$\Omega_c h^2$} & \makebox[\wB][c]{$U[0.1,\,0.15]$}
& \makebox[\wC][c]{$w_a$} & \makebox[\wD][c]{$U[-3,\,2]$} \\

\makebox[\wA][l]{$H_0$} & \makebox[\wB][c]{$U[60,\,80]$}
& \makebox[\wC][c]{$\alpha$} & \makebox[\wD][c]{$U[0,\,1.2]$} \\

\makebox[\wA][l]{$\ln(10^{10}A_s)$} & \makebox[\wB][c]{$U[3.0,\,3.1]$}
& \makebox[\wC][c]{$\beta$} & \makebox[\wD][c]{$U[0,\,0.6]$} \\

\makebox[\wA][l]{$n_s$} & \makebox[\wB][c]{$U[0.9,\,1.1]$}
& \makebox[\wC][c]{$\varphi_\mathrm{ini}$} & \makebox[\wD][c]{$U[0,\,2]$} \\

\makebox[\wA][l]{$\tau$} & \makebox[\wB][c]{$U[0.04,\,0.1]$}
& & \\
\hline\hline
\end{tabular}
\end{table}

To assess the extent to which the $\beta\varphi\mathrm{CDM}$ model is supported by current cosmological observations, we modified the \textsc{camb} code~\cite{Lewis:1999bs,Li:2023fdk,Li:2014eha,Hu:2013twa} to compute cosmology with nonminimal coupling~\cite{Li:2014eha}. At the background level, we solve the evolution equations based on Eq.~\eqref{eq:EOM_phi}, while directly using Eq.~\eqref{eq:Evo_DM} to describe the DM evolution. With this implementation, the coefficient $V_0$ in the potential and the value of the present scalar-field $\varphi_0$ must be specified in advance. We therefore employ the Broyden iteration method to ensure consistency between the input parameters and the resulting solution~\cite{Broyden:1965isd,gay1979some}. We also take into account the impact of the nonminimal coupling on linear perturbations~\cite{Li:2023fdk}. To obtain the posterior distributions of the model parameters, we use the public sampling code \textsc{Cobaya}~\cite{Torrado:2020dgo} to perform Markov chain Monte Carlo (MCMC) analyses, and \textsc{PolyChord} to perform nested sampling~\cite{Handley:2015fda,Handley:2015vkr}. Since extreme values of the nonminimal coupling parameters may cause the shooting method to fail, we impose priors on the parameter ranges, as listed in Table~\ref{tab:priors}. The sampling is terminated when the remaining evidence contained in the live points satisfies $\Delta \ln Z < 0.001$~\cite{Ashton:2022grj,Handley:2015vkr}. We use the public package \textsc{GetDist}~\cite{Lewis:2019xzd} to analyze and visualize the sampling results. The  datasets used in this work include:

\begin{itemize}
  \item CMB: the \textsc{CamSpec} likelihood based on the Planck PR4 NPIPE high-multipole ($\ell>30$) TT, TE, and EE spectra~\cite{Rosenberg:2022sdy}, together with the low-multipole temperature and E-mode polarization power spectra, $C_\ell^{TT}$ and $C_\ell^{EE}$, derived using \textsc{Commander} and \textsc{SimAll}~\cite{Planck:2019nip}, respectively. We also include the CMB lensing likelihood constructed from the combination of Planck PR4 NPIPE~\cite{Carron:2022eyg} and ACT DR6~\cite{AtacamaCosmologyTelescope:2025blo}.
  \item BAO: the DESI Y3 BAO measurements listed in Table IV of the DR2 paper~\cite{DESI:2025zgx}.
  \item SN: the Type Ia SN samples from PantheonPlus (PP)~\cite{Brout:2022vxf}, DES-Y5~\cite{DES:2024jxu}, and DES-Dovekie~\cite{DES:2025sig}.
  \item $H_0$: we further consider the local $H_0$ measurement from Riess 2020~\cite{Riess:2021jrx}. Since the DES SN data release provides the relative distance modulus $\mu$ directly, the local $H_0$ information is incorporated as a Gaussian prior on $H_0$, rather than through the absolute magnitude $M_B$, which may weaken the Bayesian evidence.
\end{itemize}
For brevity, in what follows, we use ``baseline'' to denote the combination of the latest CMB+BAO data.

\section{Cosmological Constraints}\label{sec:constraint}

\begin{table}[b]
\centering
\caption{Cosmological constraints (mean and $68\%$ CL) for $\Lambda$CDM, $w_0w_a$CDM, and the $\beta\varphi$CDM model.}
\label{tab:cosmo_params}
\renewcommand{\arraystretch}{1.4}
\begin{tabular}{l|c|c|c}
\hline\hline
Para. & $\Lambda$CDM & $w_0w_a$CDM & $\beta\varphi$CDM \\
\hline
$\Omega_b h^2$ & $0.02232^{+0.00012}_{-0.00012}$ & $0.02224^{+0.00012}_{-0.00012}$ & $0.02221^{+0.00015}_{-0.00014}$ \\
$\Omega_c h^2$ & $0.11792^{+0.00059}_{-0.00058}$ & $0.11917^{+0.00081}_{-0.00081}$ & $0.11776^{+0.00112}_{-0.00074}$ \\
$H_0$          & $68.07^{+0.26}_{-0.26}$          & $67.36^{+0.54}_{-0.55}$          & $67.76^{+0.60}_{-0.58}$ \\
$\Omega_\mathrm{m}$ & $0.3041^{+0.0029}_{-0.0034}$ & $0.3131^{+0.0053}_{-0.0053}$ & $0.3063^{+0.0057}_{-0.0059}$ \\
$S_8$ & $0.8147^{+0.0071}_{-0.0071}$ & $0.8281^{+0.0083}_{-0.0083}$ & $0.8286^{+0.0095}_{-0.0104}$ \\
$w_0$          & $\cdots$                  & $-0.808^{+0.054}_{-0.054}$        & $\cdots$ \\
$w_a$          & $\cdots$                  & $-0.724^{+0.225}_{-0.200}$        & $\cdots$ \\
$\alpha$       & $\cdots$                  & $\cdots$                  & $0.661^{+0.234}_{-0.133}$ \\
$\beta$        & $\cdots$                  & $\cdots$                  & $0.197^{+0.079}_{-0.143}$ \\
$\mathcal{A}_{\mathrm{ini}}$ & $\cdots$   & $\cdots$                  & $1.0292^{+0.0119}_{-0.0127}$ \\
\hline\hline
\end{tabular}
\end{table}

\begin{figure}[t]
    \centering
    \includegraphics[width=0.95\linewidth]{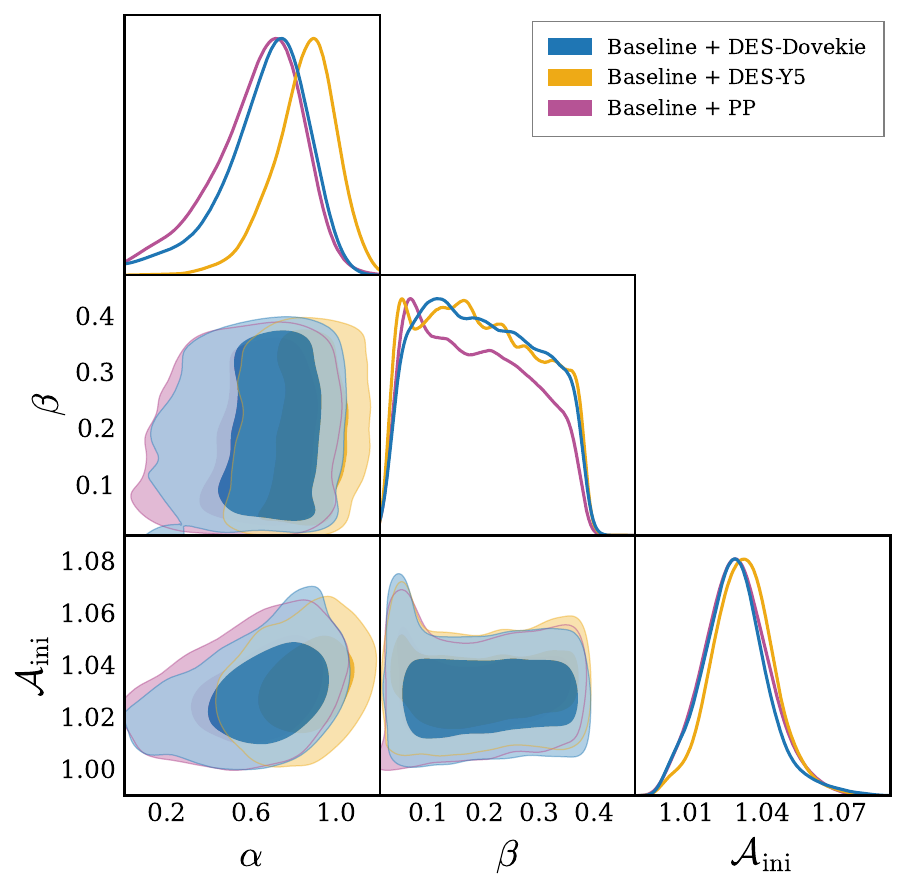}
    \caption{2-dimensional marginalized constraints on the $\beta\varphi$CDM model parameters from the combined datasets of Baseline and SN measurements from DES-Dovekie (blue), DES-Y5 (orange), and PantheonPlus (purple), respectively.}
    \label{fig:2d_params}
\end{figure}

We perform cosmological analyses using the Baseline dataset in combination with different SN samples, as shown in Fig.~\ref{fig:2d_params}, where $\mathcal{A}_\mathrm{ini}=\mathcal{A}(\varphi_\mathrm{ini})$ denotes the initial coupling strength. In all cases, the results indicate about $2\sigma$ evidence for a nonzero coupling and a nontrivial scalar potential, providing statistical support for the $\beta\varphi\mathrm{CDM}$ model. When SN data with a stronger preference for dynamical dark energy are adopted, the preferred value of $\alpha$ shifts toward larger values, while the coupling parameter $\beta$ retains a broad distribution that remains nearly unchanged across different SN samples. This suggests that the coupling in $\beta\varphi\mathrm{CDM}$ is primarily supported by the CMB and BAO data, rendering the result less sensitive to the recent debate over the DES SN datasets.

To illustrate more clearly the impact of the nonminimal coupling on cosmological evolution, we present the key cosmological parameters constrained by the latest Baseline+DES-Dovekie dataset as a representative example. Compared to the $w_0w_a\mathrm{CDM}$ model, the parameters $\{H_0,\Omega_\mathrm{m}\}$ in the $\beta\varphi\mathrm{CDM}$ model remain closer to their $\Lambda\mathrm{CDM}$ values. As we show in the next section, this smaller value of $\Omega_\mathrm{m}$ plays an important role in the apparent crossing behavior.

For different dataset combinations, the cosmological evidence for each model is quantified through the Bayesian evidence~\cite{Handley:2015vkr,Jeffreys:1939xee},
\begin{equation}
  Z_i=\int_\Omega P(D|\theta_i,M_i)P(\theta_i |M_i)\,\mathrm{d}\theta_i,
\end{equation}
with the results summarized in Table~\ref{tab:Bayes}. Notably, for all dataset combinations considered, $\beta\varphi\mathrm{CDM}$ yields the largest Bayes factor over the $w_0w_a$CDM model. In particular, for the Baseline+DES-Y5 dataset, the cosmological evidence for $\beta\varphi\mathrm{CDM}$ relative to $\Lambda\mathrm{CDM}$ reaches $\Delta \ln Z=+4.7$, indicating that $\beta\varphi\mathrm{CDM}$ can effectively describe late-time DDE behavior. In addition, when the local $H_0$ likelihood is included, the evidence for $\beta\varphi\mathrm{CDM}$ is further strengthened by an appreciable amount. We will discuss in the next section how the $\beta\varphi\mathrm{CDM}$ model helps alleviate the current cosmological discrepancies.

\begin{table}[t]
\centering
\caption{Bayesian model comparison for different dataset combinations. The numbers report the Bayes factor relative to $\Lambda {\rm CDM}$, $\Delta \ln Z\equiv \ln Z_{\rm model}-\ln Z_{\Lambda \rm CDM}$ for each dataset choice.}
\label{tab:Bayes}
\renewcommand{\arraystretch}{1.25}

\newcommand{\wA}{3.8cm}
\newcommand{\wB}{1.8cm}
\newcommand{\wC}{1.8cm}

\begin{tabular}{l|c c}
\hline\hline
\makebox[\wA][l]{Datasets (+Baseline)} & \makebox[\wB][c]{$w_0w_a\mathrm{CDM}$} & \makebox[\wC][c]{$\beta\varphi\mathrm{CDM}$} \\
\hline
\makebox[\wA][l]{DES-Dovekie} & \makebox[\wB][c]{$+0.50$} & \makebox[\wC][c]{$+1.19$} \\
\makebox[\wA][l]{DES-Dovekie+$H_0$} & \makebox[\wB][c]{$+0.24$} & \makebox[\wC][c]{$+2.60$} \\
\hline
\makebox[\wA][l]{DES-Y5} & \makebox[\wB][c]{$+3.83$} & \makebox[\wC][c]{$+4.65$} \\
\makebox[\wA][l]{DES-Y5+$H_0$} & \makebox[\wB][c]{$+0.98$} & \makebox[\wC][c]{$+3.11$} \\
\hline
\makebox[\wA][l]{PantheonPlus} & \makebox[\wB][c]{$-1.62$} & \makebox[\wC][c]{$+0.81$} \\
\makebox[\wA][l]{PantheonPlus+$H_0$} & \makebox[\wB][c]{$+1.37$} & \makebox[\wC][c]{$+4.14$} \\
\hline\hline
\end{tabular}
\end{table}

\section{Discussions}\label{sec:discussion}

In this section, we discuss how the $\beta\varphi\mathrm{CDM}$ model affects the background cosmological evolution.

\subsection{Dynamical dark energy}\label{subsec:DDE}

An advantage of the $\beta\varphi\mathrm{CDM}$ model is that it does not require tuning the quintessence mass to trigger thawing at a particular expansion rate $H(z)$, thus naturally avoiding the coincidence problem with fine-tuning. As long as the DM density remains much larger than the DE density, the quintessence continuity Eq.~\eqref{eq:EOM_phi} is dominated by the coupling term. Once the DM density decreases to be comparable to that of quintessence as the Universe expands, the scalar field naturally begins to thaw from $\varphi=0$. Therefore, the $\beta\varphi\mathrm{CDM}$ model provides a natural explanation for the dynamics of DE in the late Universe, closely linked to the onset of DE domination.

\begin{figure}[b]
    \centering
    \includegraphics[width=0.95\linewidth]{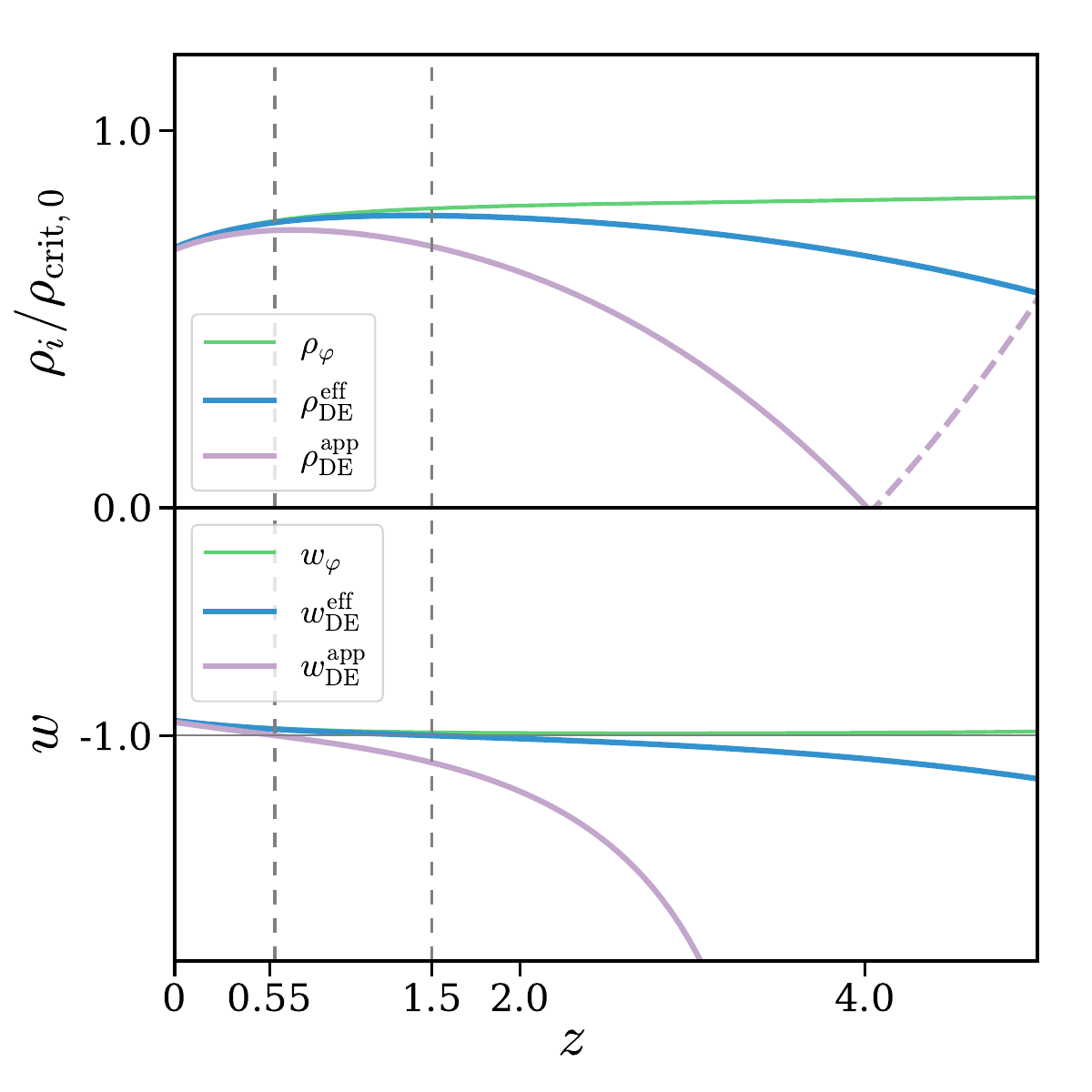}
    \caption{Low-redshift evolution of the energy densities and reconstructed equations of state for the scalar field, dark matter, the effective DE, and the apparent DE defined in Eq.~\eqref{eq:rho_DE_eff} and Eq.~\eqref{eq:rho_DE_app}. The phantom-crossing behavior in the reconstructed $w(z)$ occurs at different redshifts depending on the bookkeeping used to define the dark-energy sector.}
    \label{fig:wrho_lowz}
\end{figure}

Given the intriguing crossing behavior inferred from DESI BAO and DES SN within the CPL parametrization, it is meaningful to examine the DE evolution in the $\beta\varphi\mathrm{CDM}$ model. By construction, the quintessence accounts for all components of the dark sector other than DM, which is consistent with the definition of DE. Since the model is based on a canonical scalar field, the quintessence energy density $\rho_\varphi$ must decrease with time, implying $w_\varphi>-1$, as shown with green curves in Fig.~\ref{fig:wrho_lowz}.

However, since the nonminimal coupling causes the DM density to deviate from the standard $a^{-3}$ scaling, it is not appropriate to compare $\rho_\varphi$ directly with the DE density in the $w_0w_a\mathrm{CDM}$ model. From an observational perspective, a common procedure is to take a fitted value of $\rho_\mathrm{c}h^2$ in the $\beta\varphi\mathrm{CDM}$ model and artificially extract a DM component that evolves as $a^{-3}$. This leads to a so-called effective DE density $\rho_\mathrm{DE}^\mathrm{eff}$ and a misestimate of DM density $\Delta\rho_\mathrm{DM}$~\cite{Chakraborty:2025syu, Das:2005yj},
\begin{align}
  \rho_{\rm DE}^{\rm eff}&=\rho_\varphi +\Delta\rho_{\rm DM},\label{eq:rho_DE_eff}\\
  \Delta\rho_{\rm DM} &\equiv \rho_{\rm DM} - \rho_{{\rm DM},0}a^{-3}.
  \label{eq:delta_rho_dm}
\end{align}
Based on the definition of the EoS parameter, $w=-\dot{\rho}/(3H\rho)-1$, such an effective DE density yields
\begin{equation}
  w_\mathrm{DE}^\mathrm{eff}=\frac{w_\varphi}{1+\Delta\rho_\mathrm{DM}/\rho_\varphi}.\label{eq:w_DE_eff}
\end{equation}
As shown in Fig.~\ref{fig:phi_evolution}, ${\cal A}(\varphi)$ increases in the late Universe, corresponding to an energy transfer from DE to DM, and hence $\Delta \rho_{\rm DM}<0$. It is this negative density correction that causes $w_{\rm DE}^{\rm eff}$ to cross $w=-1$ in nonminimally coupled models. A similar crossing behavior can also be achieved in quintom models by introducing a noncanonical field~\cite{Feng:2004ad,Feng:2004ff,Guo:2004fq}. To illustrate this effect, we adopt representative parameter values $\{\alpha,\beta,\varphi_{\rm ini}\}=\{0.8, 0.2, 0.6\}$ based on the results of Table~\ref{tab:cosmo_params}, and then compute $w_{\rm DE}^{\rm eff}$, as shown with blue curves in Fig.~\ref{fig:wrho_lowz}. We find that $w_{\rm DE}^{\rm eff}$ indeed crosses $w=-1$ near $z=1.5$.

\begin{figure*}[t]
    \centering
    \includegraphics[width=0.9\linewidth]{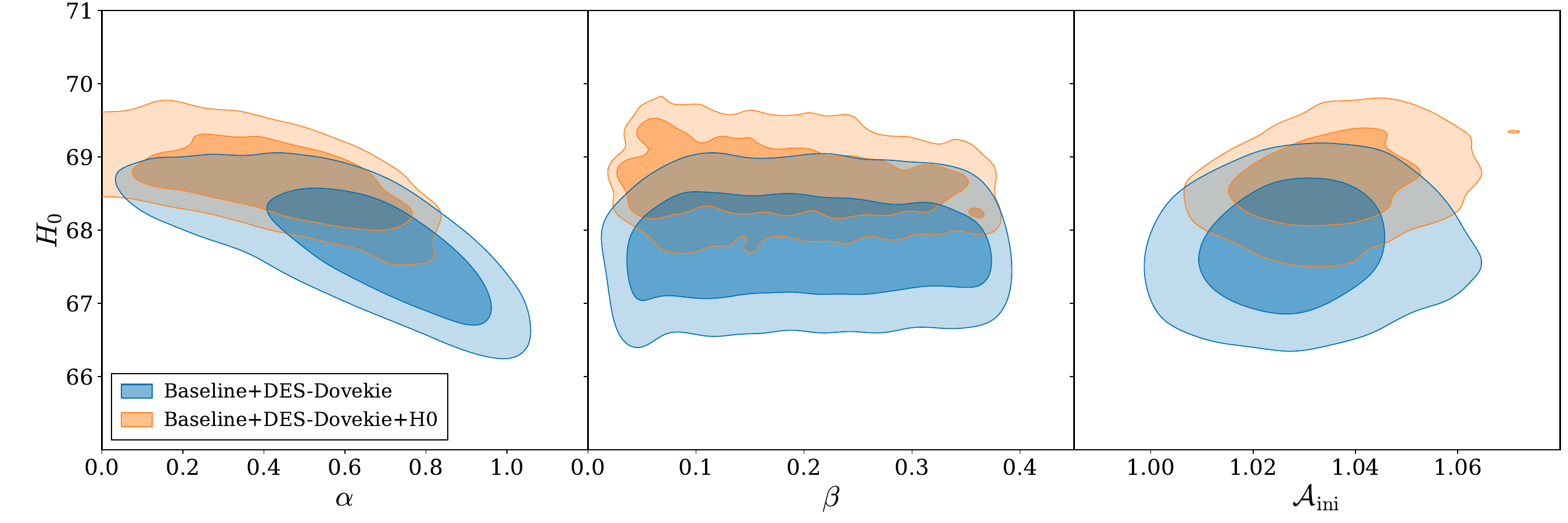}
    \caption{Cosmological constraints on $H_0$ and three model parameters {$\alpha,\beta,\mathcal{A}_\mathrm{ini}$} in the $\beta\varphi\mathrm{CDM}$ model from Baseline+DES-Dovekie (blue) and Baseline+DES-Dovekie+$H_0$ (orange) datasets. }
    \label{fig:H0}
\end{figure*}

The crossing behavior exhibited by the effective DE is appealing, but not entirely satisfactory, since the corresponding crossing redshift is higher than the value $z\sim 0.5$ suggested by DESI~\cite{DESI:2025zgx, DESI:2025fii}. However, it should be noted that the crossing behavior is a phenomenological result inferred within the CPL parametrization for the apparent DE component actually seen by the $w_0w_a\mathrm{CDM}$ model. Therefore, the DM component to be subtracted should be $\rho_{\mathrm{DM},w_0w_a\mathrm{CDM}}$, rather than $\rho_\mathrm{DM}$ inferred within $\beta\varphi\mathrm{CDM}$ as listed in Table~\ref{tab:cosmo_params}. This subtlety was first pointed out in Ref.~\cite{Wang:2025znm} for a nontrivial example with opposite trends between the effective and apparent DE components, as we will define shortly below.

To distinguish it from the effective DE defined above, we refer to the DE component constructed in the above way as the apparent DE associated with the $w_0w_a$CDM model, with the density and EoS parameters given by~\cite{Wang:2025znm}
\begin{align}
  \rho_\mathrm{DE}^\mathrm{app}&\equiv \left(\rho_\varphi+\rho_\mathrm{DM}\right)_{\beta\varphi\mathrm{CDM}}-(\rho_\mathrm{CDM})_{w_0w_a\mathrm{CDM}},\label{eq:rho_DE_app}\\
  w_\mathrm{DE}^\mathrm{app}&=\frac{w_\mathrm{DE}^\mathrm{eff}}{1+(\rho_\mathrm{DM,0}-\rho_\mathrm{CDM,0}^\mathrm{CPL})a^{-3}/\rho_\mathrm{DE}^\mathrm{eff}},\label{eq:w_DE_app}
\end{align}
Due to the energy transfer from DM to DE in the early Universe, $\rho_\mathrm{DM}$ in $\beta\varphi\mathrm{CDM}$ is slightly smaller than that in $w_0w_a\mathrm{CDM}$. As a result, the crossing behavior of this apparent DE, which is directly relevant for model comparison, appears at a lower redshift. As shown in Fig.~\ref{fig:wrho_lowz}, the EoS of the apparent DE crosses $w=-1$ at $z=0.56$, in better agreement with the DESI result obtained within the $w_0w_a\mathrm{CDM}$ framework.

In this section, we have examined separately the evolution of $\rho_\varphi$, $\rho_{\rm DE}^{\rm eff}$, and $\rho_{\rm DE}^{\rm app}$. A potentially confusing point is which of them should be identified with the actual DE evolution inferred from observations. Recall that DE is usually defined as the part of the dark sector remaining after removing DM. Within the $\beta\varphi\mathrm{CDM}$ framework, the nonstandard evolution of $\rho_{\rm DM}$ is explicitly given by Eq.~\eqref{eq:Evo_DM}, and the remaining dark-sector component is entirely quintessence, corresponding to $\rho_\varphi$. This definition is theoretically complete and does not require introducing any additional prescription in parameter fitting. However, for interacting DE-DM models, the true evolution of DM cannot be determined. This implies that, when comparing DE behavior across different models, the DM evolution must first be aligned. Since the study of the crossing behavior is in practice a comparison to the $w_0w_a\mathrm{CDM}$ model, one should isolate from $\beta\varphi\mathrm{CDM}$ the DM component seen by the $w_0w_a\mathrm{CDM}$ model, which yields $\rho_{\rm DE}^{\rm app}$ in Eq.~\eqref{eq:rho_DE_app}. In addition, the previously widely used definition $\rho_{\rm DE}^{\rm eff}$ enforces $\Delta\rho_{\rm DM}=0$ at $z=0$, presuming all DM today are cold, which is yet to be confirmed observationally. 
In summary, $\rho_\varphi$ is the theoretical definition of quintessence within the $\beta\varphi\mathrm{CDM}$ model, while $\rho_{\rm DE}^{\rm app}$ corresponds to the observational DE relevant for model comparison in studies of crossing behavior.

\subsection{$H_0$ tension}\label{subsec:H0tension}

As shown in Fig.~\ref{fig:phi_evolution}, the evolution of the scalar field in the $\beta\varphi\mathrm{CDM}$ model also undergoes a transition around the epoch of matter-radiation equality ($z_\mathrm{eq}$). Through the DM-DE coupling, this might lead to a reduction in the DM mass and allow the dark sector to exhibit an evolution analogous to that of early DE (EDE)~\cite{Niedermann:2020dwg,Poulin:2018cxd,Lee:2022cyh,Gomez-Valent:2022bku}. Furthermore, as shown in Table~\ref{tab:Bayes}, the Bayes evidence for $\beta\varphi\mathrm{CDM}$ is significantly improved when the local $H_0$ likelihood is included. Similar effects have also been discussed in Ref.~\cite{Pitrou:2023swx,Uzan:2023dsk}. Therefore, it is worthwhile to examine the potential of the $\beta\varphi\mathrm{CDM}$ model to alleviate the $H_0$ tension.

We analyze the posterior distribution of $H_0$ and its correlation with the main model parameters, as shown in Fig.~\ref{fig:H0}. The results indicate that only the potential parameter $\alpha$ exhibits a weak correlation with $H_0$. Such a correlation is expected since the late-time evolution of DE can significantly alter the CMB angular-diameter distance. However, $\beta$ and $\mathcal{A}_\mathrm{ini}$ have almost no impact on $H_0$, which largely rules out the possibility that the nonminimal coupling acts effectively as EDE. In addition, without including the local $H_0$ likelihood, the posterior of $H_0$ constrained within $\beta\varphi\mathrm{CDM}$ remains almost unchanged relative to that in $\Lambda\mathrm{CDM}$. These results suggest that the $\beta\varphi\mathrm{CDM}$ model cannot resolve the $H_0$ tension, at least within the currently allowed parameter space. The increase in Bayes evidence appears to arise mainly because the additional model parameters weaken the constraining power of the data on $H_0$.

The reason for this limitation is that the scalar field $\varphi$ in the current $\beta\varphi\mathrm{CDM}$ model cannot decrease rapidly after $z_\mathrm{eq}$. As a result, its ability to alleviate the $H_0$ tension is strongly constrained by the CMB power spectra. A special form of the coupling, capable of driving a rapid thawing of $\varphi$, may provide a way to overcome this difficulty and is therefore worth investigating in future work.

Extending the nonminimal coupling beyond the dark sector to the SM particles, or even to gravity, may help alleviate the cosmological difficulties associated with the $H_0$ tension. In Ref.~\cite{Ye:2024zpk}, Ye pointed out that a quintessence nonminimally coupled to gravity can potentially resolve the $H_0$ tension. However, such modified gravity models may affect the observable Universe and therefore typically require additional new physics to evade the stringent laboratory constraints on the fifth forces~\cite{Qin:2022qnk,Landim:2024wzi,GRAVITY:2025ahf}. By contrast, in the $\beta\varphi\mathrm{CDM}$ model, the nonminimal coupling is confined entirely to the dark sector and is thus subject to far fewer constraints~\cite{Carroll:2008ub,Bai:2009it,Carroll:2009dw}. The bound derived from the tidal tails, $\beta<0.5$ for the $\beta\varphi\mathrm{CDM}$ model, already covers nearly the entire posterior parameter space~\cite{Kesden:2006vz}.

\section{Conclusions}\label{sec:conclusion}

A sign-switching interaction is naturally realized in the $\beta\varphi\mathrm{CDM}$ model through nonminimally coupled quintessence, in which the late-time dynamics of dark energy are naturally explained by the transition of the dominant component of the Universe. Current cosmological data provide evidence at about the $2\sigma$ level for both the nonminimal coupling and the nontrivial scalar potential. Compared with both $\Lambda\mathrm{CDM}$ and $w_0w_a\mathrm{CDM}$, the $\beta\varphi\mathrm{CDM}$ model receives stronger statistical support from current cosmological observations with or without $H_0$ prior. From the perspective of model mismatch, we point out that DE in $\beta\varphi\mathrm{CDM}$ can also exhibit a crossing behavior when mapped onto the $w_0w_a\mathrm{CDM}$ framework. More refined forms of the coupling and the potential may yield even stronger cosmological support and may also provide a promising route toward alleviating the $H_0$ tension.

\renewcommand{\acknowledgmentsname}{Acknowledgments}
\begin{acknowledgments}
We are grateful to Tian-Nuo Li, Meng-Xiang Lin, and Gen Ye for insightful discussions. 
This work is supported by the National Key Research and Development Program of China Grants No. 2021YFC2203004, No. 2021YFA0718304, and No. 2020YFC2201501, 
the National SKA Program of China (Grants Nos. 2022SKA0110200 and 2022SKA0110203),
the National Natural Science Foundation of China Grants No. 12422502, No. 12547110, No. 12588101, No. 12235019, No. 12447101, No. 12473001, No. 11975072, No. 11835009, No. 11805031, and No. 12533001,
and the China Manned Space Program Grant No. CMS-CSST-2025-A01 and No. CMS-CSST-2025-A02.
We also acknowledge the use of the HPC Cluster of ITP-CAS.
\end{acknowledgments}

\section*{Data Availability}
The data that support the findings of this article are not publicly available because they are owned by a third party and the terms of use prevent public distribution. The data are available from the authors upon reasonable request.

\appendix

\section{The attractor behavior of $\varphi$}\label{app:attractor}

In this Appendix, we summarize the analytic origin of the freezing behavior during radiation domination and of the attraction toward the vicinity of $\varphi=0$ during matter domination. For the model adopted in the main text,
\begin{equation}
    \mathcal{A}(\varphi)=1+\frac{1}{2}\beta\varphi^2,
    \qquad
    V(\varphi)=V_0 e^{\alpha\varphi},
\end{equation}
with $\alpha>0$, $\beta>0$, and $V_0>0$, Eq.~\eqref{eq:Evo_DM} can be written as
\begin{equation}
    \rho_\mathrm{DM}=\bar{\rho}_\mathrm{DM}(a)\mathcal{A}(\varphi),
    \qquad
    \bar{\rho}_\mathrm{DM}(a)\propto a^{-3}.
\end{equation}
Therefore, the effective force acting on the scalar field becomes
\begin{equation}
    V_{\mathrm{eff},\varphi}
    =V_{,\varphi}+\frac{\mathcal{A}_{,\varphi}}{\mathcal{A}}\rho_\mathrm{DM}
    =\alpha V_0e^{\alpha\varphi}+\beta\bar{\rho}_\mathrm{DM}\varphi.
    \label{eq:app_effective_force}
\end{equation}
During radiation domination, $\rho_r\propto a^{-4}$ dominates over both $V$ and $\bar{\rho}_\mathrm{DM}\propto a^{-3}$. Hence $V_{\mathrm{eff},\varphi}/H^2\ll1$, and the K--G equation reduces approximately to $\ddot{\varphi}+3H\dot{\varphi}\simeq0$. The scalar velocity is damped as $\dot{\varphi}\propto a^{-3}$, so the field remains frozen near its initial value.

During matter domination, the coupling-induced term in Eq.~\eqref{eq:app_effective_force} becomes dynamically important. The instantaneous minimum of the effective potential is defined by
\begin{equation}
    \alpha V_0e^{\alpha\varphi_\mathrm{min}}
    +\beta\bar{\rho}_\mathrm{DM}\varphi_\mathrm{min}=0 .
    \label{eq:app_minimum_condition}
\end{equation}
Thus, at finite $\bar{\rho}_\mathrm{DM}$, $\varphi=0$ is not the exact minimum because $V_{\mathrm{eff},\varphi}(0)=\alpha V_0$. Instead, Eq.~\eqref{eq:app_minimum_condition} gives
\begin{equation}
    \varphi_\mathrm{min}
    =-\frac{1}{\alpha}
    W\left(\frac{\alpha^2V_0}{\beta\bar{\rho}_\mathrm{DM}}\right),
\end{equation}
where $W$ is the Lambert $W$ function. In the matter era, it still holds $\bar{\rho}_\mathrm{DM}\gg V_0$, so
\begin{equation}
    \varphi_\mathrm{min}
    \simeq
    -\frac{\alpha V_0}{\beta\bar{\rho}_\mathrm{DM}} .
    \label{eq:app_minimum_approx}
\end{equation}
Thus the minimum is slightly shifted to the negative side of $\varphi=0$, with a displacement suppressed by $V_0/\bar{\rho}_\mathrm{DM}$. In this sense, $\varphi=0$ should be understood as the limiting attractor of the matter era, approached in the limit $\bar{\rho}_\mathrm{DM}\gg V_0$, rather than as the exact finite-density minimum.
The uniqueness and stability of this minimum follow from
\begin{equation}
    V_{\mathrm{eff},\varphi\varphi}
    =\alpha^2V_0e^{\alpha\varphi}+\beta\bar{\rho}_\mathrm{DM}>0 ,
\end{equation}
for $\alpha>0$, $\beta>0$, $V_0>0$, and $\bar{\rho}_\mathrm{DM}>0$. The effective potential is therefore strictly convex and has a unique stable minimum, toward which deviations are damped by Hubble friction.

This attractor behavior is not tied to every detail of the functional forms in Eq.~\eqref{eq:AV_choice}. What is essential is that the coupling sector has a stable even expansion around $\varphi=0$, so that $\mathcal{A}_{,\varphi}/\mathcal{A}$ is approximately linear in $\varphi$, while the bare potential supplies only a subdominant force during matter domination. Under these conditions the coupling-induced term drives the field toward a stable minimum close to $\varphi=0$. However, the precise location and uniqueness of the minimum do depend on the chosen forms of $\mathcal{A}(\varphi)$ and $V(\varphi)$; in the present model they are guaranteed by the convexity shown above.

\bibliography{refer}

\end{document}